\documentstyle[aps,multicol,prl,epsf]{revtex}
\tighten
\draft

\begin{document}

\title{Transport in Coupled Quantum Dots: Kondo Effect Versus Anti-Ferromagnetic Correlation}
\author{C.A. B\"usser$^1$, E.V. Anda$^1$, A.L Lima$^1$, Maria A. Davidovich$^1$, 
and G. Chiappe$^2$ \\
\it $^1$Departamento de F\'{\i}sica, Pontificia Universidade 
Cat\'olica do Rio de Janeiro, C.P. 38071-970, Rio de Janeiro, RJ, Brazil \\
\it $^2$Departamento de F\'{\i}sica,
Facultad de Ciencias Exactas y Naturales, \it Universidad de Buenos Aires
Ciudad Universitaria, 1428, Buenos Aires, Argentina}
\maketitle

\begin{abstract}
The interplay between the Kondo effect and the inter-dot magnetic interaction in a coupled-dot system is studied. An exact result for the transport properties at zero temperature is obtained by diagonalizing a cluster, composed by the double-dot and its vicinity, which is connected to leads. It is shown that the system goes continuously from the Kondo regime to an anti-ferromagnetic state as the inter-dot interaction is increased. The conductance, the charge at the dots and the spin-spin correlation are obtained as a function of the gate potential. 
\end{abstract}

\begin{multicols}{2}

In the last years  electron transport through a quantum dot (QD) has been the 
subject of many experimental  and theoretical 
\cite{variosT,varios T1,variosE,Alfredo,Goldhaber,Georges} 
investigations. The possibility  of measuring the Kondo effect, the basis of heavy-fermion physics, in a system with a QD has been proposed by several theoretical studies\cite{variosT,Alfredo}. 
Its detection is a difficult task since it depends on 
several different energy scales and their relative sizes, such as the coupling 
constant between the QD and leads, the Kondo temperature $T_K$ and the energy 
spacing between the dot levels, which implies small size dots. Such a very small QD was recently obtained and the Kondo effect was measured \cite{Goldhaber}.
This confirms the possibility of studying many of the properties 
of strongly correlated metals and insulators in artificially
constructed mesoscopic structures. The advantage of these systems is that their parameters can be continuously changed by modifying the applied external potentials so that different regimes can be studied. 

The physics associated to heavy-fermion compounds in the vicinity of what 
is called the quantum phase transition is signed by the   competition between 
the spin correlation among the magnetic atoms and between the atoms and the conduction electrons \cite{Coqblin}. 
The latter gives rise to the Kondo effect and the 
former tends to 
create a ferro or anti-ferromagnetic ground state destroying the Kondo regime. 
The study of the two impurity Kondo Hamiltonian was proposed to clarify this 
important problem, as it includes in the model the two interactions responsible for this competing behavior \cite{Jones}. 
The competition gives rise to several phenomena such as  
magnetic impurity correlation, one and two-stage Kondo effect 
and, according to the particle-hole symmetry, Fermi and non-Fermi 
liquid behavior \cite{siete}.

There have been extensive experimental transport studies in the double-dot system
\cite{ocho}.
They were mainly designed to analyze the double-dot molecule.
From the theoretical view point transport through a double-dot  has
received a considerable attention mainly restricted to the study 
of high temperature, Coulomb blockade phenomenon \cite{nueve}.

At very low temperature, the study of a completely different physics, 
created by the interplay 
of the Kondo effect and the inter-dot anti-ferromagnetic correlation, is now feasible due to the recent possibility of constructing 
very small dots. 
An interesting analysis of these phenomena in a two-QD system, based on qualitative arguments and 
slave-boson mean-field theory, has recently appeared \cite{Georges}.
The behavior of the conductance was studied for a system where the double occupancy at 
the dots was eliminated from the Hilbert space. 
Although this method 
becomes exact  when the number of spin degrees of freedom is infinite, some caution is necessary when applying it to spin 1/2 systems.

The purpose of this letter is to present a numerically 
exact calculation of the transport properties of a double-dot system to investigate the competition between the Kondo effect and the anti-ferromagnetic correlation. The parameters are chosen to reflect real experimental conditions and, in particular, the intra-dot Coulomb repulsion is taken to be finite. 
The conductance is obtained for several inter-dot coupling constants. 
The density of states, the charge inside the dots and the various 
spin-spin correlation functions are calculated in order to characterize the state of the system.

The very low temperature properties ($T\ll T_K$ ) of the system are obtained by using the Lanczos method \cite{Dagotto} to calculate the ground state of a small cluster containing the two dots, which is embedded into the leads. The conductance calculated according to this procedure tends very rapidly to its exact result as the size of the cluster is increased.  From this view point our calculation is 
numerically exact since we were able to reach convergence, within 1\% error, for a small cluster. For $T>T_K$ we use a 
self-consistent solution for the equation of motion for the one-particle Green function \cite{Hubbard}.
This approximation eliminates all the low lying excitations in the vicinity of 
the Fermi level and, as a consequence, the Kondo effect. However,  it provides an adequate description of the Coulomb blockade high temperature regime\cite{varios T1}.

The system is represented by an Anderson two-impurity first-neighbor tight-binding Hamiltonian,
\begin{equation}
H = H_c + t \sum_{\sigma\atop ij} c_{i\sigma}^+ c_{j\sigma}
\label{H}
\end{equation}
where $t$ is the nearest-neighbor hopping in the leads.
The Hamiltonian $H_c$ for the cluster of $2M+2$ sites containing the two dots denoted by $\alpha$ and $\beta$ and the other sites numbered from $1$ to $M$ and $\bar 1$ to $\bar M$ can be written as  
\begin{eqnarray}
H_c &=& V \sum_{\sigma\atop r=\alpha,\beta}  n_{r \sigma} +
\frac{U}{2} \sum_{r \sigma} n_{r \sigma} n_{r \bar\sigma} +
t'' \sum_{\sigma} (c^+_{\alpha \sigma} c_{\beta \sigma} + c.c.)  
\nonumber \\ & &
 + t' \sum_\sigma (c^+_{\alpha \sigma} c_{\bar 1 \sigma} +  
% \nonumber \\ & &
c^+_{\beta \sigma}
 c_{ 1 \sigma} + c.c.) + t \sum_{\sigma\atop i,j} c^+_{i \sigma} 
c_{j \sigma}
\label{Hc}
\end{eqnarray}
where $U$ and $V$ represent, respectively, the electronic repulsion 
and the gate potential at the dots, $t''$ is the inter-dot interaction, $t'$ 
the coupling between the dots and the leads, and the sub-indices $i$ and $j$,  in the last term, run over the sites of the cluster other than the dot sites.
We restrict our analysis to the case of two identical dots.  
An inter-dot magnetic exchange term $J$ is not explicitly included in the Hamiltonian since it is not an independent parameter but a function of the inter-dot tunneling ($J \sim t''^2/U$)\cite{Coqblin}.

To describe the very low temperature physics of the system, we calculate the one particle Green functions 
$G_{\alpha \beta}$ at 
the dots. They are imposed to satisfy a Dyson equation  $\hat G
= \hat g + \hat g \hat T \hat G$ where  $\hat g$ is the cluster Green 
function  matrix and $\hat T$ is the matrix of the coupling Hamiltonian  
between the cluster  and the rest of the system. The undressed 
Green function $ \hat g $ 
is calculated using the cluster ground 
state obtained by the Lanczos method. The Dyson equation proposed is equivalent to the chain 
approximation in a kinetic  energy diagrammatic 
expansion for the Hubbard Hamiltonian \cite{Metzner}. To be consistent, the charge of the dressed and undressed cluster is imposed to be the same. We calculate 
$ \hat g$ as a combination of the Green functions of $n$  and $n+1$ 
electrons with weights $(1-p)$ and $p$, $\hat g=(1-p) \hat g_n+p \hat g_{n+1}$ \cite{Valeria}. 
In this case the charge of the undressed cluster is 
$q_c=(1-p)n+p(n+1)$. The charge of the cluster 
linked to the leads can be expressed as $Q_c = 2\int_{-\infty}
^{\epsilon_f} \sum_i \mbox{Im} G_{ii}(\omega) d\omega$ where $i$ runs through 
all the cluster sites and the factor 2 includes the electronic spin. 
This equation together with the  
imposed condition $q_c = Q_c$ constitute a system of two equations, which requires a self-consistent solution  to obtain $p$ and $n$ as a function of the parameters of the system. Due 
to the symmetry of the problem, it is useful to represent the Hamiltonian using 
the two-dot bonding and anti-bonding states. In this case, according to the Keldysh formalism\cite{Keldysh},
the conductance can be written as 
$ \sigma = e^2/h [t'^2 \rho(\epsilon_f)]^2 \vert G_d^+ -G_d^- \vert ^2$ 
  where $G_d^\pm = 1 / (\omega - \epsilon^\pm - \Sigma_c(w) - \Sigma_d^\pm(w))$ are the one particle 
Green  functions at the dots represented in the bonding and anti-bonding wave functions, with energies $\epsilon^\pm = V \pm t'$. $\Sigma_c(w)$ and $\Sigma_d^\pm(w)$  are the self-energies corresponding to the leads and to the many-body interaction, respectively.

The expression for the current is obtained  
by considering explicitly that the system is a Fermi liquid 
$\lim_{\omega  \to 0} \mbox{Im} \Sigma_d^\pm(\omega)\propto\omega^2$, 
which implies that there are no dissipative  
processes taking place at the Fermi level. It is important to point out that our 
diagonalization self-consistent procedure satisfies, without any imposition, 
the  Luttinger-Ward identity 
$\mbox{Im} \int_{-\infty}^{\epsilon_f} ({\partial \Sigma_d^\pm(\omega)}/
{\partial \omega}) G_d^\pm(\omega) d\omega = 0$ \cite{Luttinger}.
This ensures the fullfilment of the Fridel Sum Rule and of the Fermi liquid properties.

We  have as well studied this problem 
through  the equation of motion for the $G^+$ and $G^-$ using a 
self-consistent decoupling 
procedure, proposed by Hubbard \cite{Hubbard}.
Since it eliminates all the low lying excitations 
involved in the Kondo effect and properly  describes the high energy states  
we suppose it to be an adequate  solution for high temperatures, 
$T > T_k$. It has 
been used with success  to study the  Coulomb blockade behavior of the 
current 
going through a highly confined region, as in a QD\cite{varios T1}. Within this  
approximation the many-body self-energy can be  expressed as  
$\Sigma_d^\pm(w) = {n^\pm U (\omega - \epsilon_{\pm})}/
{(\omega - i\eta -\epsilon^\pm - (1-n^\pm) U)}$. It requires the 
self-consistent calculation of the charge at the bonding and anti-bonding
state $n^\pm$.   
As shown below the two treatments are equivalent 
in the limit of strong anti-ferromagnetic  
inter-dot coupling regime, where there is no Kondo ground state.

In order to provide a complete characterization of the ground 
state we have compared the inter-dot  
spin-correlation $\langle\vec S_{d\alpha} \vec S_{d\beta}\rangle$ with the spin-correlation of one dot and its nearest-neighbor  
conduction electrons $\langle\vec S_{c1} \vec S_{d\beta}\rangle$. For the sake of simplicity, we have calculated these objects on the 
cluster ground state.

\begin{figure}
\epsfxsize=8.7cm 
\epsfysize=10.5cm 
\centerline{\epsfbox{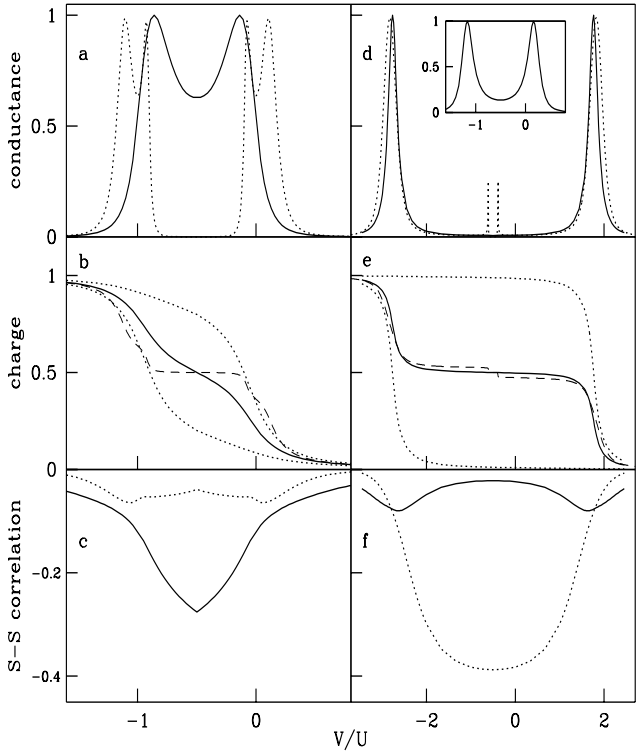}}
\narrowtext
\caption{Conductance, charge and spin correlation with 
$\Gamma/U=0.08$, for weak ($t''/U= 0.16 $) and strong ($t''/U= 2.0$) 
coupling. Conductance in units of $e^2/h $: weak (a) 
and strong (d) coupling; 
exact-cluster (continuous) and Hubbard approximation (dotted). 
The inset in (d) corresponds to the intermediate 
coupling ($t''/U=0.4$) in the exact-cluster approximation. 
Charge: weak (b) and strong (e) coupling; exact-cluster results for 
charge per spin in each dot(continuous) and in the $+ -$ states (dotted); 
Hubbard approximation results for charge per spin in each dot (dashed). Spin correlation: 
weak (c) and strong (f) coupling; dot-conduction-electron, 
$\langle \vec S_{c1} \vec S_{d\beta}\rangle$, (continuous) and dot-dot, 
$\langle\vec S_{d\alpha} \vec S_{d\beta}\rangle$, (dotted).}
\label{fig:cond1}
\end{figure}

The transport properties 
are studied for weak and strong inter-dot coupling. 
We solve an eight-atom cluster with the parameter 
$\Gamma/U=t'^2/WU = 0.08$, where $W$ is the leads bandwidth. This value 
corresponds approximately to 
the device where the Kondo effect has recently been measured
\cite{Goldhaber}
and is compatible with first principle calculated parameters \cite{Molinari}. 
The current, the dots charge and the spin-spin correlation for  weak 
inter-dot coupling $(t''/U=0.16)$, as a function of $V$, are presented at the left part of Fig. \ref{fig:cond1}. 
For $V>0.2$ the dot levels are above the Fermi level and   
there is no current. As $V$ is reduced, the two 
dots become partially charged because the anti-bonding state enters 
into resonance. When the total charge at the dots is $N\sim1$ there is a peak in the conductance which reaches the value of $e^2/h$. As soon as the system enters into the fluctuating valence regime, corresponding to $\epsilon^-$ below the Fermi energy, the Kondo peak appears creating a new channel for the electrons to flow. These features are shown in Figs.~1a and 1b. The spin correlation $\langle\vec S_{c1} \vec S_{d\beta}\rangle$, continuous line in Fig.~1c, increases. As $V$ is further augmented extra charge enters into the dots 
and the conductance diminishes. As
$\epsilon^-$ goes far beyond the 
Fermi energy the Kondo temperature reduces exponentially so that $ t''^2/(U T_K)$ increases, partially quenching the Kondo effect. The conductance has a minimum for $V/U = -0.5$, when 
the system has electron-hole symmetry and $N\sim 2$.
However, the current is large since the system is still within the  
Kondo regime, as reflected by the relationship 
 $\langle \vec S_{d\alpha} \vec S_{d\beta}\rangle \ll \langle \vec S_{c1} \vec S_{d\beta}\rangle$, shown in Fig.1c.
 The large conductance in the inter-peak region 
is due to the Abrikosov-Suhl resonance created by the Kondo effect. This channel is absent when $T>T_K$. The Hubbard 
solution, also shown in Fig.~1a, can 
be taken to be a high temperature description of the same problem. It preserves the structure of the Coulomb blockade, with no 
conductance between the peaks, which are splitted  into two due to the inter-dot interaction.

It is important to notice the different way the charge enters into the dots in  the Coulomb blockade and in the Kondo regimes. In the former, the charge presents a plateau due to electronic repulsion while in the latter, the Kondo peak going through the Fermi level permits a continuous entrance of the charge. This behavior is shown in Fig.~1b.

Finally in the case of the strong inter-dot interaction $(t''/U=2)$,  
shown at the right part of 
Fig.~1, the Kondo regime is almost quenched by the anti-ferromagnetic coupling. The Kondo spin-spin 
correlation, although small, has its maximum in the near  vicinity 
of the conductance peaks, where the Kondo temperature is high, while in the whole intermediate region anti-ferromagnetism  
controls the conduction. 
In this case, the conductance results obtained by the two approaches, shown in Fig.~1d, are almost identical, reflecting the fact that the system is outside the Kondo regime for most values of the  gate potential. In the region between the peaks the Hubbard approximation presents a very small contribution to the conductance, due to the inter-dot interaction splitting, which is absent in the exact-cluster solution.  

When the dot-dot interaction has an intermediate value $(t''/U= 0.4)$ the conductance in the region between the two peaks results to be slowly dependent upon the gate potential, as shown in the inset of Fig.~1d. This behavior has been characterized as a two-plateau structure in the
conductance as a function of the gate voltage by a mean-field slave-boson 
approximation\cite{Georges}, which gets a strictly zero conductance in the vicinity of the gate potential value for which the system has electron-hole symmetry.
However we do not get neither a strict plateau structure nor a complete  quenching  of the Kondo effect. In the inter-peak region, for finite Coulomb repulsion $U$ the current is not zero, although it is small when the dots are strongly coupled. The dependence of the conductance for the electron-hole symmetric situation as a function of the inter-dot coupling is shown in Fig.~2a.
It is clearly seen that the interplay between the Kondo effect and the anti-ferromagnetic correlation does not produce any abrupt transition. Our results show a crossover behavior as the inter-dot interaction is varied.

\begin{figure}
\epsfxsize=3.5cm 
\epsfysize=3.5cm
%\centerline
{\epsfbox{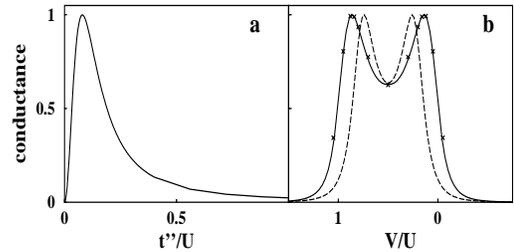}}
\caption{ Conductance in units of $e^2/h$ (a) as a function of the inter-dot coupling for $V/U=-0.5$, (b) as a function of gate potential for cluster of 4 sites(dashed), 8 sites(continuous), 12 sites(crosses) }
\label{fig:CONV}
\end{figure}

In order to study the convergence of our results with the size of the cluster we present in Fig.~2b the weak coupling limit conductance for three cluster sizes. A comparison of the conductance for eight and twelve-site clusters, which differs
at most by 1\%, allows us to conclude that our results are numerically exact. 
\begin{figure}
\epsfxsize=7.5cm 
\epsfysize=8cm
\centerline{\epsfbox{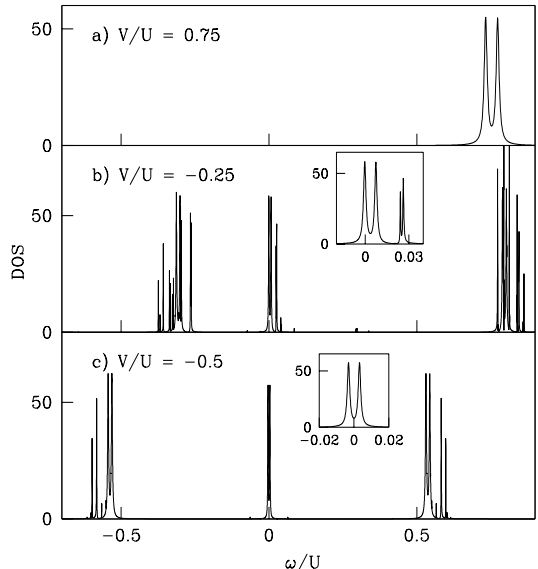}}
\caption{ DOS  for weak coupling regime and three values of gate 
potential $V$. $\epsilon_f=0$, $\Gamma/U=0.05$ and  $t''/U = 0.02$. Insets show detail of Kondo peak. } \label{fig:DOS}
\end{figure}

The analysis of the density of states (DOS) for various values of the gate 
potential helps to clarify the different circumstances involved. In 
Fig.~3a the DOS shows a relatively large and broad 
resonant peak above the Fermi energy which is splitted by the inter-dot interaction. It is the DOS of a one-body problem since the dots are empty, $N\sim0$.
The case where $N\sim1$ and the conductance has its maximum value $e^2/h$ is presented in Fig.~3b. There is an 
asymmetric Kondo peak much 
thinner than the two resonant levels which are separated by the Coulomb 
interaction $U$. The inset in the figure shows the  Fermi level located at one 
of the Abrikosov-Suhl sub-peaks. Finally Fig.~3c shows the 
electron-hole symmetric condition, $N\sim2$, where the Kondo resonance is just 
in the middle of the two Coulomb blockade peaks. The Fermi energy coincides 
with the position of the pseudo-gap of the Kondo peak, as shown in the inset. 
These results show that in the last two cases the system is in the Kondo regime
and explain the reduction of the conductance for the situation corresponding to Fig.~3c relative to the one of Fig.~3b. 

In summary, we have obtained a numerically exact $T=0$ solution for the transport properties of a coupled two-dot system. We show that the interplay between the interactions that give rise to the Kondo effect and the anti-ferromagnetic coupling are explicitly reflected on the conductance of the system. The different regimes have been characterized through a detailed study of the charge at the dots and the spin-spin correlations.

We acknowledge the agencies CNPq, CAPES, FAPERJ, Antorchas/Vitae/Andes grant A-13562/1-3,CONICET and Fundacion Antorchas for financial support.

\end{multicols}

\end{document}